\documentclass{aa}
\usepackage{amsmath}
\usepackage{psfig}
\usepackage{rotate}
\usepackage{astrobib}
\usepackage{journals}
\newcommand{\figwidth}{0.55\textwidth}
\begin{document}

\title{A jet model for the broadband spectrum of XTE J1118+480}
\subtitle{Synchrotron emission from radio to X-rays in the Low/Hard spectral state}

\author{Sera Markoff\inst{
1}\fnmsep\thanks{Humboldt research fellow}
\and Heino Falcke\inst{1} \and Rob Fender\inst{2}}
\institute{Max-Planck-Institut f\"ur Radioastronomie, Auf dem H\"ugel
69, D-53121 Bonn,
Germany 
\and Astronomical Institute `Anton Pannekoek' and Center for High
Energy Astrophysics, University of Amsterdam, 
Kruislaan 403, 1098 SJ Amsterdam, The Netherlands}

\titlerunning{Jet model for XTE J1118+480}
\authorrunning{Markoff et al.}
\offprints{smarkoff@mpifr-bonn.mpg.de}

\date{Accepted for publication in Astronomy \& Astrophysics Letters}

\abstract{Observations have revealed strong evidence for powerful jets
in the Low/Hard states of black hole candidate X-ray binaries.
Correlations, both temporal and spectral, between the radio --
infrared and X-ray bands suggest that jet synchrotron as well as
inverse Compton emission could also be significantly contributing at
higher frequencies.  We show here that, for reasonable assumptions
about the jet physical parameters, the broadband spectrum from radio
through X-rays can be almost entirely fit by synchrotron emission. We
explore a relatively simple model for a relativistic, adiabatically
expanding jet combined with a truncated thermal disk conjoined by an
ADAF, in the context of the recently discovered black hole binary XTE
J1118+480.  In particular, the X-ray power-law emission can be
explained as optically thin synchrotron emission from a shock
acceleration region in the innermost part of the jet, with a cutoff
determined by cooling losses. For synchrotron cooling-limited particle
acceleration, the spectral cutoff is a function only of dimensionless
plasma parameters and thus should be around a ``canonical'' value for
sources with similar plasma properties.  It is therefore possible that
non-thermal jet emission is important for XTE J1118+480 and possibly
other X-ray binaries in the Low/Hard state.\keywords{X-rays: binaries
-- X-rays: individual: XTE J1118+480 -- radiation mechanisms:
non-thermal -- stars: winds, outflows --black hole physics --
accretion, accretion disks} } \maketitle

\section{Introduction}

Observations are providing increasing evidence that black hole
candidate (BHC) X-ray binaries (XRBs) produce powerful collimated
outflows when in the Low/Hard X-ray state (LHS).  This state is
characterized by a nonthermal power-law in the X-ray band and little,
if any, thermal disk contribution (e.g.,
\citeNP{Nowak1995,Poutanen1998}). At least three systems in the LHS
(Cyg X-1, 1E 1740.7-2942 and GRS 1758-258) have directly resolved
radio jets on scales from AU to parsecs.  However, XRB jets also
reveal themselves in the broadband LHS spectra with a flat-to-inverted
radio synchrotron spectrum, analogous to the signature emission of
jets in compact radio cores of AGN
\cite{BlandfordKoenigl1979,HjellmingJohnston1988,FalckeBiermann1999}.
Furthermore, new data suggest a continuation of this radio synchrotron
emission to much higher frequencies.  The spatial, spectral and
temporal evidence for powerful jets from XRB BHCs in the LHS are
compiled and discussed in \citeN{Fender2001}.

We know that jets play a significant role in the emission of Active
Galactic Nuclei (AGN), even dominating the spectrum from radio through
TeV $\gamma$-rays in the case of BL Lacs, with emission
from optically thin synchrotron up to even $100$ keV and higher
(e.g. \citeNP{Pianetal1998}). By analogy, if the flat,  optically
thick synchrotron spectrum in XRBs, commonly attributed to jets,
indeed extends into the NIR and optical regimes \cite{Fender2001}, one
would expect a corresponding optically thin power-law from shock
acceleration at even higher frequencies. This is irrespective of
whether the jet emission is boosted or not, since optically thick and
optically thin emission would have similar Doppler factors.  Shock
acceleration is likely to be present in XRBs, given that optically
thin power-law spectra are observed during their radio outbursts
(e.g., \citeNP{FenderKuulkers2001}, and refs. therein).

Still, the majority of current models for the broadband (X-ray)
spectra of BHC XRBs focus only on the contribution of thermal disk
plus coronal inverse Compton (IC) emission (for a review, see
\citeNP{Poutanen1998}) and ignore any jet contribution, even though
jets can be an integral part of X-ray binaries and their systems
(e.g., \citeNP{EikenberryMatthewsMorgan1998,MirabelRodriguez1999,Fender2001}).

As an example for the possible importance of jet emission, we use the
recently discovered XRB XTE J1118+480
\cite{Remillardetal2000}, which has been observed in the
radio through X-rays (see \citeNP{Hynesetal2000}, hereafter H00,
\citeNP{Fenderetal2001}, hereafter F01, and references therein), and
is also at high enough Galactic latitude to allow the first ever EUV
detections of an X-ray transient (H00).  The system is likely a BHC in
the LHS (H00;\citeNP{RevnivtsevSunyaevBorozdin2000,Woodetal2000}).
Although jets were not directly resolved with MERLIN to a limit of
$<65(d/{\rm kpc})$ AU (at 5 GHz; F01), its radio emission shows the
flat characteristic jet spectrum.

\section{Basic Model and Estimates}

We consider that an accretion disk is responsible for the
extreme-ultraviolet (EUV) data and contributes in the optical (H00;
\citeNP{Garciaetal2000}).  However, the power-law seen in the 
X-ray spectrum indicates that an additional component must be
present. So far, this has been presumed to arise from the
Comptonization of some `seed' photons by a hot, thermal corona
(\citeNP{Poutanen1998}).

One commonly invoked physical explanation for the LHS is that a
standard thin, optically thick disk \cite{ShakuraSunyaev1973} exists
only down to some transition radius $r_{\rm tr}\sim10^2-10^3 r_{\rm
s}$, ($r_{\rm s}=2GM_{\rm bh}/c^2$), where the flow becomes hot and
non-radiative
(e.g., \citeNP{Liuetal1999,Esinetal2001}).  We do
not, however, include an entire disk model here, since there are
several models already in existence exploring this issue (see above,
and review in \citeNP{Poutanen1998}). Instead, we include a
representative thermal spectrum for the inner edge of the cool disk
both in the direct emission, in our calculations of cooling rates in
the jet and as seed photons for IC.

With this approach the temperature, $T_{\rm d}$, and luminosity,
$L_{\rm d}$, for this inner edge can be roughly determined by fitting
a black body to the EUV data.  As discussed in H00, uncertainty in the
local absorption leads to large variations in the possible EUV flux
for XTE J1118+480.  Here, we take the highest absorption value
presented, $N_{\rm H} = 1.15\cdot10^{20}$ cm$^{-2}$, because it
provides a solid upper limit to any thermal disk contribution (see
Fig.~1).

Taking the distance and BH mass to be $1.8$ kpc and $6 M_\odot$,
respectively \cite{McClintocketal2001}, the fit to the EUV data gives
$T_{\rm d}\sim1.5\cdot 10^5$ K and $L_{\rm d}
\sim 5 \cdot 10^{35}$ erg s$^{-1}$.  For  an annulus with scale
width $\sim r_{\rm tr}$, we find as an order of magnitude estimate
$r_{\rm tr}\approx900 r_{\rm s}$, in agreement with current models
(e.g., \citeNP{Liuetal1999}). The luminosity of a
standard accretion disk is dominated by the inner edge and its
radiative efficiency is $q_{\rm l}=\frac{1}{4} r_{\rm s}/r_{\rm tr}$
(\citeNP{FrankKingRaine1992}, Eq.~5.20), yielding a rough estimate for
the accretion rate of $\dot M_{\rm d}\sim q_{\rm l}^{-1} L_{\rm
d}/c^2\simeq3\cdot 10^{-8}M_\odot$ yr$^{-1}$.  Hence, in the following
we use a reference value of $\dot M_{\rm d}=\dot
m_{-7.5}\;3\times10^{-8} M_\odot$ yr$^{-1}$.  Within $r_{\rm tr}$ we
consider a hot, ADAF-like flow which does not significantly contribute
to the spectrum.

For accreting black holes it has been argued that the jet power is of
order $Q_{\rm jet}\sim q_{\rm j}\dot M_{\rm d} c^2$ with an
efficiency inferred to be of order $q_{\rm j}=10^{-1}-10^{-3}$
\cite{FalckeBiermann1999}. However, $q_{\rm j}$ is
essentially a free parameter, with $q_{\rm j}\ll1$, and we define
$q_{\rm j,-2}=q_{\rm j}/10^{-2}$.  While the jet formation itself is
very difficult to model, the physics of calculating most of the jet
emission is relatively straightforward because the flat-to-inverted
spectrum stems from the part of the jet where it is basically
undergoing free expansion.  Here we build on the jet emission model
outlined in \citeN{FalckeMarkoff2000}, and references therein.

At the inner edge of the hot accretion flow, plasma is ejected out
from symmetric nozzles, where it becomes supersonic.  The jets then
accelerate along the axes through their pressure gradients up to bulk
Lorentz factors $\gamma_{\rm j}\simeq2-3$, and expand sideways with
their initial proper sound speed $\gamma_{\rm s}\beta_{\rm
s}c\simeq0.4c$ for a hot electron/proton plasma. This implies low
relativistic Mach numbers around ${\cal M}\sim5$. The velocity field,
density and magnetic field gradient then come naturally from the Euler
equation (see, e.g., \citeNP{Falcke1996}).  The dependencies of the
magnetic field $B$ and density $n$ on distance are then similar to,
but slightly stronger than, the canonical $r^{-1}$ and $r^{-2}$
dependencies for conical jets, respectively
\cite{BlandfordKoenigl1979,HjellmingJohnston1988,FalckeBiermann1995}.

In this way, the basic physical properties governing the emission at
each point in the jet are fixed after specifying the jet power, and
the initial conditions at the nozzle.  We make the simplification of
assuming a maximal jet, which follows from the Bernoulli equation
(\citeNP{FalckeBiermann1995}) when the internal energy, here dominated
by the magnetic field, is equal to the bulk kinetic energy of
particles. This is consistent with a magnetic launching mechanism. The
plasma is assumed to originate in the hot accretion flow and therefore
contains equal numbers of protons and electrons, with hot electrons at
a temperature approaching $T_{\rm e}=T_{\rm e,10}\cdot 10^{10}$ K in
various ADAF models (e.g., \citeNP{Manmoto2000}). The electron Lorentz
factor of the peak will be at $\gamma_{\rm e}\sim4\cdot T_{\rm e,10}$.

In AGN jets, the high frequency, optically-thin power-laws are taken
to be the result of synchrotron emission from particles being
shock-accelerated along the jet (e.g.,
\citeNP{MarscherGear1985}).  In such a case the crucial parameter
for the high energy emission is the location $z_{\rm acc}$ of the
first particle acceleration region in the jet.

Near the shock region in each jet, $B^2(r)/8\pi\simeq0.25q_{\rm j}
\dot M_{\rm d} c^2/(c\gamma_{\rm j}\beta_{\rm j} \pi r^{2})$, yielding
a reference value of $B\simeq2 \cdot 10^6\,{\rm G}\cdot
\sqrt{q_{j,-2}\dot m_{-7.5}/\gamma_{\rm j}\beta_{\rm j}} (r/10r_{\rm
s})^{-1}$ for the parameters discussed above. Similarly, the particle
density is given by $n\simeq 10^{14}\,{\rm cm}^{-3}\cdot (q_{j,-2}\dot
m_{-7.5}/\gamma_{\rm j}\beta_{\rm j}) (r/10r_{\rm s})^{-2}$.

Once the plasma, assumed to be injected at the base of the jet with a
Maxwellian distribution, reaches the shock region, the standard
diffusive shock acceleration process redistributes the particles
into a power-law, starting roughly at $\gamma_{\rm e,min}\sim4 T_{\rm
e,10}$.  The initially injected distribution has an index
$p\simeq1.5-2$ (resulting from a relativistic shock, see, e.g.,
\citeNP{HeavensDrury1988}\footnote{Because the plasma is mildly
relativistic at the shock region, the temperature-dependent adiabatic
index will be closer to 4/3 rather than the non-relativistic value of
5/3.  This yields a spectral index of $1.5\la p \le 2$, where $p=2$ is
for non-relativistic shock-acceleration.})  steepening
by $\sim 1$ due to increased cooling above a break energy $E_{\rm
b}\approx (\tau_{\rm r} \beta_{\rm cool})^{-1}$, where $\tau_{\rm r}$
is the residence time of the plasma in the acceleration region, and
$\beta_{\rm cool}$ are the constants giving the energy dependence of
the cooling, which for synchrotron and Compton losses go as
$\dot{E}\approx-\beta_{\rm cool} E^2$.  This steepened spectrum has
the index of $p\simeq2.5-3$, typically found in the optically thin
synchrotron emission of both AGN and X-ray binaries.  For the case of
XTE J1118+480, the unbroken X-ray power-law (see Fig.~1), implies
$p\simeq2.6$ for $E>E_{\rm b}$.  The acceleration ceases when
the particles reach the energy $E_{\rm e, max}=\gamma_{e,max}m_{\rm
e}c^2$ where the cooling/loss rates equal that of acceleration.  These
rates are dependent both on the energy of the particle, as well as the
local physical parameters.

The shock acceleration rate is given as
\begin{equation}
t_{\rm
acc}^{-1}=\frac{3}{4}\left(\frac{u_{\rm
sh}}{c}\right)^2\frac{eB}{m_{\rm e}c \xi
\gamma_{\rm e}},
\end{equation}
where $u_{\rm sh}$ is the shock speed in the plasma frame. The
parameter $\xi< c \beta_e/u_{\rm sh}$ \cite{Jokipii1987} is the ratio
between the diffusive scattering mean free path and the gyroradius of
the particle, and has a lower limit at $\xi=1$. We account for energy
losses at the shock via adiabatic losses, particle escape, IC and
synchrotron. In our case the latter dominates and we have

\begin{equation}
t_{\rm syn}^{-1}=\frac{4}{3}\sigma_{\rm T} \gamma_{\rm e}
\beta_{\rm e}^2
\frac{U_{\rm B}}{m_{\rm e} c}
\end{equation}
where $\sigma_{\rm T}$ is the Thomson cross-section and $U_{\rm B}$ is
the energy density of the magnetic field.  The maximum energy is then
found by solving $t_{\rm acc}^{-1}=t_{\rm syn}^{-1}$, yielding
$\gamma_{\rm e,max}\sim10^8\,\left(\xi
B\right)^{-0.5}\left(\frac{u_{\rm sh}}{c}\right)$.  If we define as a
reference value $\xi=\xi_2 100$, the maximum synchrotron frequency is
\begin{equation}
\nu_{\rm max}=0.29 \nu_{\rm c} \simeq 1.2\cdot 10^{20} \xi_2^{-1}
\left(\frac{u_{\rm sh}}{c}\right)^2 \;\;\; {\rm Hz}
\end{equation}
where $\nu_{\rm c}\simeq\frac{3}{4\pi} \gamma_{\rm e,max}^2
(eB)/(m_{\rm e} c)$ is the critical synchrotron frequency. This
maximum corresponds approximately to the rollover of the power-law
cutoff, and for $u_{\rm sh}\sim\beta_{\rm s}c$, we find a cutoff of
$\sim80$ keV.  This cutoff is not dependent on the magnetic field, the
jet power, or the shock location as long as we are in the synchrotron
cooling dominated regime.  Because we would expect XRBs to have
similar shock structures, once $\xi$ is roughly fixed
observationally---thus determining the scattering between magnetic
irregularities in the diffusive shock process---we should get similar
cutoffs for different sources and accretion rates.

The location of the initial shock acceleration region is set by the
frequency where the flat, highly self-absorbed synchrotron spectrum
turns over into the optically thin power-law produced at the shock.
From back-extrapolating the X-ray power-law, we can see that this
maximum self-absorption frequency has to be somewhere in the
IR/optical regime at $\sim 10^{14.5}$ Hz.  For the parameters
discussed here this would be at a distance of roughly $z_{\rm
acc}\simeq50r_{\rm s}$, with a jet radius of $r\simeq10r_{\rm s}$, and
the magnetic field and particle densities derived above.  The total
synchrotron luminosity, integrated up to the highest energies, is a
few times $10^{36}$ erg ${\rm s}^{-1}$, which is $\sim10\%$ of the
total jet power ($Q_{\rm j}\sim q_{\rm j,-2}\dot
m_{-7.5}\,2\cdot10^{37}$ erg s$^{-1}$).

Some fraction of X-rays created close to the nucleus by the jet will
of course either directly impinge on, or be scattered by hot electrons
into, the cold disk, resulting in a reflection component.  However,
since this is very dependent on the disk geometry, and because the
feature is very weak or absent for XTE J1118+480
\cite{McClintocketal2001b}, we do not calculate this process here.

\section{Results of Numerical Modeling}
To obtain a more detailed broadband spectrum we have used the full
numerical calculations for a jet model as described in
\citeN{FalckeMarkoff2000}, with the addition of a particle
acceleration region. The model takes into account the relativistic
Doppler shifts, adiabatic losses, electron acceleration and loss
timescales at the acceleration region as described above, and
integrates the synchrotron and IC emission along the jet.  The
resulting fit is shown in Figure 1 with the parameters given in the
caption.

The relevant free parameters are the inclination angle $\theta_{\rm
i}$, the shock distance $z_{\rm acc}$, the scattering ratio $\xi$, and
the jet power parameterized by $q_{\rm j}$.  While the electron
temperature $T_{\rm e}$ (or $\gamma_{\rm e,min}$) and the fraction of
thermal particles accelerated can also be adjusted, they are not
independent of $q_{\rm j}$ -- decreasing these parameters will decrease
the radiative efficiency and increase $q_{\rm j}$.  Similarly, a more
realistic disk model (e.g., \citeNP{Esinetal2001}) could yield a lower
$r_{\rm tr}$ and $\dot M$, also increasing
$q_{\rm j}$.  Because we require charge balance between protons and
electrons for the jet plasma, and a low $T_{\rm e}$ in line with ADAF
models, the particles are in sub-equipartition with the magnetic
field.

The flat spectrum from radio to optical is due to the optically thick
synchrotron emission from the jet at $z>z_{\rm acc}$, while the X-ray
power-law is the optically thin synchrotron emission dominated by the
shock acceleration region (at $z\sim z_{\rm acc}$).  For the
conditions in the jet, $E_{\rm b}$ lies very close to the peak of the
thermal electron distribution, and we only see the steepened
power-law.  Because of the large ratio between magnetic field and
photon densities, IC emission does not play a significant role for the
parameters chosen here (see Fig.~1). Synchrotron and IC emission from
the pre-shock region ($z<z_{\rm acc}$) could in principle show up in
the soft and very hard X-rays respectively as a function of $T_{\rm
e}$, but the EUV points place an effective upper limit $T_{\rm
e}\la2\cdot10^{10}$ K, and so for this source these features are
rather weak.

The jet length may be constrained by the turnover seen in the $\sim
2-15$ GHz range, which gives a length of  $\sim 4\cdot10^{13}$ cm,
i.e. $\sim1.5$ mas (at 1.8 kpc), in the GHz range with a $\nu^{-0.9}$
scaling of the size \cite{Falcke1996}.

The optically thin synchrotron power-law component in the X-rays
depends on the existence of a diffusive shock acceleration region. The
current limit on $\gamma_{\rm e}$ comes from synchrotron losses
instead of IC, due to the weak disk emission, which results in a low
external photon density.  In this case, the location of the X-ray
cutoff is determined by $\xi$ and is fit observationally, keeping in
mind that $\xi$ is typically under a few $10^2$ (e.g.,
\citeNP{Jokipii1987}).  A ``canonical'' cutoff at $\sim100$ keV (e.g.,
\citeNP{Poutanen1998}) corresponds to $\xi\sim100$, and $\nu_{\rm
max}\propto\xi^{-1}$.  We also plot both $\xi=10$ and $\xi=100$ in
Figure 1, in order to illustrate how the cutoff could theoretically
venture into the realm of future high-energy missions like INTEGRAL
with energies $\la$ a few MeV.  For XTE J1118+480, such a high cutoff
may be necessary, as \citeN{McClintocketal2001b} see no cutoff below
$\sim160$ keV.

\begin{figure}
\centerline{\psfig{figure=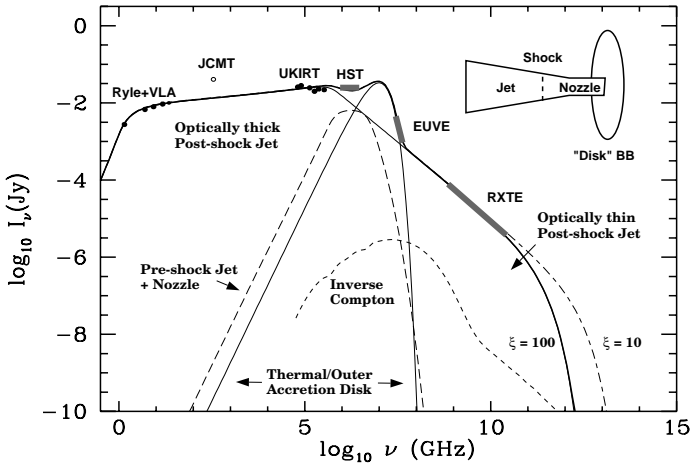,width=\figwidth}}
\caption[]{\label{fig} Fit to data from H00, with the
exception of the VLA radio points (Hjellming et al., private comm.),
the Ryle Telescope and a non-simultaneous JCMT point (see F01, and
refs. therein).  Included is a schematic indicating the different
emission components.  Parameters are $M_{\rm bh}=6 M_\odot$, $r_{\rm
nozz}=1.5 r_{\rm s}$, $T_{\rm d}=1.7\cdot10^5$ K, $L_{\rm
d}=5.25\cdot10^{35}$ ergs s$^{-1}$, $\dot m=0.1$, $L_{\rm
j}=2.6\cdot10^{36}$ erg s$^{-1}$, $r_{\rm tr}=747 r_{\rm s}$, $q_{\rm
j,-2}=1.5$, $\xi_2=1$, $z_{\rm acc}=45 r_{\rm s}$,
$\theta_i=32^\circ$.  The fraction of thermal particles accelerated at
the shock was fixed at 50\%. The dot-dashed line shows the cutoff for the
case of $\xi_2=0.1$}
\end{figure}

\section{Conclusion}
{}From our modeling we conclude that in the LHS of XTE J1118+480 a
significant jet contribution from radio through IR is present, and
with reasonable assumptions and a small efficiency could extend even
up to the hard X-rays.  We cannot exclude that XTE J1118+480, with its
rather low luminosity, is a special source. However, our model is
scalable and thus likely applicable to other XRBs in the LHS, when a
flat radio spectrum is present.

An important element of the model is that the inner ADAF-like flow is
very hot and radiatively inefficient, injecting (mildly) relativistic
electrons into the jet and allowing high electron energies because of
low IC cooling.  With a more dominant disk contribution, e.g., in the
High/Soft state, the jet spectrum would also change and possibly
disappear due to increased cooling.  Interestingly, in the case of a
low disk contribution, synchrotron-dominated cooling provides a
natural ``canonical'' cutoff around 100 keV, which only depends on the
dimensionless plasma parameters $\xi$ and $u_{\rm sh}/c$.  Small
variations in the spectral index of the synchrotron power-law could
explain the observed time lags in some sources
\cite{KotovChurazovGilfanov2001}.  The jet model for XTE J1118+480 may
therefore provide an interesting new perspective for the modeling of
XRBs in general.

\begin{acknowledgements}

We thank P.L. Biermann, M. Nowak, T.  Beckert and the referee for
useful discussions and comments.  We are very appreciative of data
made available to us by R. Hjellming shortly before his untimely
death.
\end{acknowledgements}

\bibliography{aamnemonic,refs}
\bibliographystyle{aa}

\end{document}